\documentclass[10pt,letterpaper,twocolumn]{article} %% two column, final layout

\usepackage{ol2}
\usepackage[colorlinks=true,urlcolor=blue, linkcolor=blue]{hyperref}
\usepackage{amsmath}

\begin{document}
\twocolumn[
 
\title{Quasiphasematched concurrent nonlinearities in periodically poled  $\bf KTiOPO_4$ for quantum computing over the optical frequency comb}
\author{Matthew Pysher,$^1$ Alon Bahabad,$^2$ Peng Peng,$^1$ Ady Arie,$^3$ and Olivier Pfister$^{1,*}$}
\address{$^1$ Department of Physics, University of Virginia, 382 McCormick Road, Charlottesville, VA 22904-4714, USA\\
$^2$ Department of Physics and JILA, University of Colorado at Boulder and NIST, Boulder, Colorado 80309, USA\\
$^3$ Department of Physical Electronics, Fleischman Faculty of Engineering, Tel Aviv University, Ramat Aviv 69978, Israel\\
$^*$Corresponding author: opfister@virginia.edu}

\begin{abstract}
We report the successful design and experimental implementation of three coincident nonlinear interactions, namely ZZZ (``type-0''), ZYY (type-I), and YYZ/YZY (type-II) second harmonic generation of 780 nm light from a 1560 nm pump beam in a single, multigrating,  periodically poled $\rm KTiOPO_4$ crystal.  The resulting nonlinear medium is the key component for making a scalable quantum computer over the optical frequency comb of a single optical parametric oscillator.   
\end{abstract}
\maketitle
]
Quantum computing is an exciting field driven by the promise of exponential speedup of \emph{a priori} arduous computational processes such as integer factoring \cite{Shor1994,Nielsen2000}.  Most  proposals for experimentally implementing quantum computing call for the use of two-state quantum systems, or qubits \cite{Braunstein2001}.  However, a quantum computer could very well use continuous quantum variables  \cite{Lloyd1999,Menicucci2006}, such as position and momentum, or the quadrature amplitude operators of the quantized field \cite{Braunstein2005a,Braunstein2003a,Cerf2007}. Recently, some of us have proposed a new and extremely scalable method for building a quantum register by use of the set of quantum harmonic oscillators (``qumodes'') defined by a single optical resonator \cite{Menicucci2008,Flammia2009}. In this proposal, the quantum correlations (entanglement) necessary for quantum computing will be implemented by a nonlinear medium placed inside the cavity, thereby realizing a sophisticated optical parametric oscillator (OPO). The sophistication stems from the fact that three different second-order nonlinear interactions must be simultaneously phasematched over the same set of cavity modes, i.e.\ must be concurrent. These interactions are parametric downconversion ($\lambda/2\mapsto\lambda$) of ZZZ (``type-0''), ZYY (type-I), and YYZ/YZY (type-II), where the first letter denotes the polarization of the pump field and the last two letters denote the polarization of the signal (entangled) beams. In previous work \cite{Pooser2005}, we demonstrated the simultaneous quasi-phasematching (QPM) of this set of interactions at room temperature for $\lambda=1490$ nm in periodically poled $\rm KTiOPO_4$ (PPKTP) with a \emph{single} period of $45.65\ \mu$m. This was a serendipitous discovery that relied upon a weak seventh-order QPM of the ZYY interaction (even though the final signal turned out to be much larger). Despite this result, designing concurrent nonlinear interactions remained difficult because the precision on the Sellmeier coefficients, as well known as they are, was still not high enough, in particular for $n_Y$. 

In this Letter, we use Fourier engineering \cite{Fejer1992, Lifshitz2005}  to achieve and demonstrate a concurrent design with \emph{low-order}, hence efficient, QPM at $\lambda=1560$ nm, close to the loss minimum of silica optical fibers. Recent advances in squeezing at and around this wavelength also make it a reasonable choice \cite{Feng2008, Mehmet2009a}. This 1560 nm design required the use of three different poling periods. Two early iterations used published Sellmeier equations \cite{Fan1987,Kato2002,Emanueli2003}. In these initial versions, the ZZZ and ZYY QPM peaks overlapped well at 1560 nm at room temperature but the YZY interaction was quasi-phasematched for 1560 nm between average temperatures of $248.7^\circ$C (for a designed phase mismatch of $\rm 1.398 \times 10^5\ m^{-1}$ at room temperature) and $300.1^\circ$C (for a designed phase mismatch of $\rm 1.410 \times 10^5\ m^{-1}$ at room temperature). From these two measurements, and considering the corrections owing to the temperature expansion of the crystal \cite{Emanueli2003}, we deduced that the phase mismatch value of the YZY process shifts with temperature with a slope of $\rm 22.34\ m^{-1}/K$. This enabled us to predict the expected phase mismatch of the YZY interaction at  $40^\circ$C to be $\rm 1.348 \times 10^5\ m^{-1}$. However, owing to the uncertainty of this linear slope correction, we adopted a multi-section design for the crystal. The 10 mm $\times$ 6mm $\times$ 1 mm crystal was divided into two sections, lengthwise.

The first section, of length 5mm, was a Fourier-engineered ZZZ/ZYY concurrence grating created using the generalized dual grid method \cite{Lifshitz2005} which was previously shown to create nonlinear photonic quasicrystals whose reciprocal lattices contain an arbitrary set of desired wave vectors \cite{Bahabad2007,Bahabad2008}. With corresponding mismatch values of $\Delta k_{ZZZ}=2.510 \times 10^5\ \rm  m^{-1}$ and $\Delta k_{ZYY}=9.061\times 10^5\rm\ m^{-1}$, we designed a quasiperiodic structure with reciprocal base vectors $k_1=\Delta k_{ZZZ}+\Delta k_{ZYY}$ and $k_2=\Delta k_{ZYY}$ such that the desired orders for phase matching the two processes are $(1,-1)$ and $(0,1)$ in this basis. 
Feeding these values into the algorithm of the dual grid method, we got the two tiling vectors of the quasiperiodic structure to be of length $3.37\ \mu m$  and $2.64\ \mu m$. The duty cycles used for the two building blocks of the structure were $0\%$ and $100\%$ respectively. This means that the $3.37\ \mu m$  building block is fabricated with a positive value of the nonlinear susceptibility, and the $2.64\ \mu m$ building block with a negative value.  The Fourier coefficients given by this structure for the ZZZ and ZYY processes are $0.112$ and $0.3855$ respectively. The reciprocal basis vectors and the duty cycles of the building blocks were chosen to both maximize the Fourier coefficients and to make the product of the Fourier coefficient and the material nonlinear coupling coefficient of the two processes approximately the same. We refer the interested reader to a detailed account of using the dual grid method for the design of quasiperiodic nonlinear photonic crystals able to phase match several different processes simultaneously \cite{Bahabad2007}.     

The second section of the crystal was composed of five parallel, 1 mm wide, gratings, of respective periods $45.9$, $46.3$, $46.7$, $47.2$, and $47.7\ \mu$m in an attempt to correctly sample the wider range of QPM variation for the YZY interaction. These periods are centered around the interpolated phase mismatch value of $\rm 1.348 \times 10^5\ m^{-1}=2\pi /46.6\ \mu$m that was obtained from the measurements with the two previous samples. The ZZZ/ZYY QPM section was as wide as the crystal and overlapped with all YZY channels. 

\begin{figure}[htb]
\begin{center}
\includegraphics[width=3.25in]{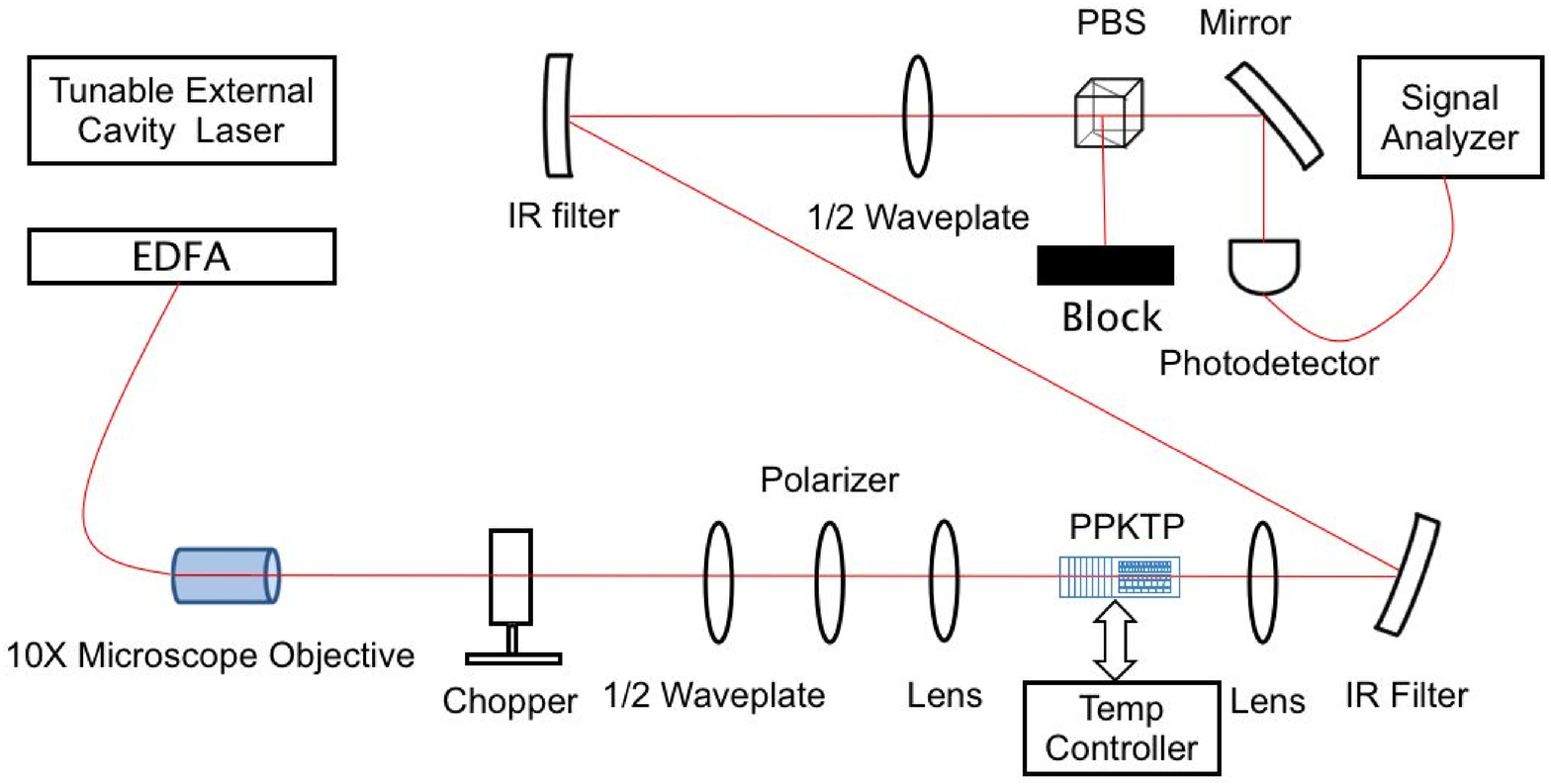}
\end{center}
%\vskip -.5in
\caption{Experimental setup. }
\label{setup}
\end{figure}

The experimental study used second-harmonic generation (SHG) with the setup shown in Fig.~\ref{setup}. 
%------------
%------------
The input 1560 nm beam was emitted by a tunable fiber laser, amplified by an erbium-doped fiber amplifier (EDFA), and then collimated and sent through a chopper wheel that allowed us to easily observe the SHG signal, amplitude-modulated at 450 Hz, on a fast-Fourier-transform signal analyzer.  After the chopper, the beam was sent through a half waveplate and polarizer, which allowed us to precisely control the polarization of the input beam.  The input beam was then focused to a waist radius of approximately 30 $\mu$m in the crystal, which was temperature controlled to the nearest hundredth of a degree.  Upon exiting the crystal, the input fundamental beam was filtered out by a pair of long-pass filters that reflected 99\% of light in the 715--900 nm wavelength range while passing over 85\% of light between 985 and 2000 nm.  Before reaching the detector, the SHG light passed through a half-waveplate and polarizing beam splitter combination, which allowed us to choose the SHG polarization to be detected.  Any residual fundamental light was filtered by the very low detection efficiency of our silicon photodiode at that wavelength. 
%------------
\begin{figure}[htb]
\begin{center}
\includegraphics[width=3.25in]{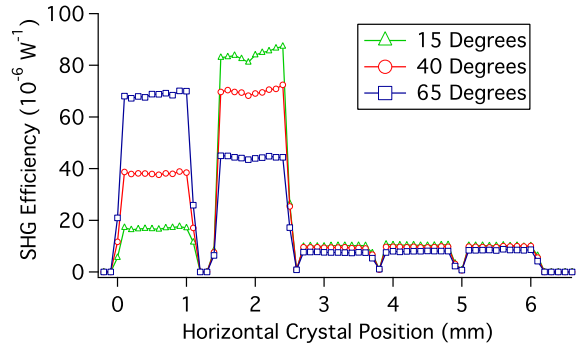}
\end{center}
\vskip -.2in
\caption{Temperature dependence of the YZY SHG signal for each of the 5 channels. The poling periods used from left to right were 45.9, 46.3, 46.7, 47.2, and 47.7 $\mu$m.}
%\vskip -.1in
\label{5channel}
\end{figure} 
%------------
The detected light was measured by taking the average of ten measurements on the signal analyzer and recording the signal at 450 Hz. The efficiency of the various nonlinear interactions was controlled by adjusting both the crystal temperature and the wavelength of the input beam.  The desired YZY poling period fell in between the 45.9 and 46.3 $\mu$m periods that were used to create our first two YZY channels.  The other three YZY channels did not yield a significant SHG signal within the temperature range obtainable by our thermoelectric controller.  Figure \ref{5channel} shows the temperature dependence of the YZY SHG signal for each of the five YZY channels.    

%------------
\begin{figure}[htb]
\begin{center}
\includegraphics[width=3.25in]{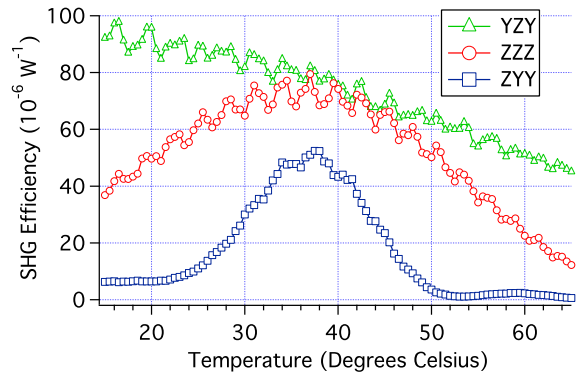}
\end{center}
\vskip -.2in
\caption{Triply concurrent SHG  as a function of temperature at 1560 nm in the $46.3\ \mu$m-period YZY channel. }
\label{tripconctempdata}
\end{figure} 
%------------
It appears that the $45.9\ \mu$m period corresponds to a QPM temperature larger than $65^\circ$C (which was the limit of our oven), while the $46.3\ \mu$m period is optimized just below $15^\circ$C. 
Despite the fact that none of our YZY channels used the exact poling period needed to put the SHG peak at 1560 nm at 40 degrees, the large temperature acceptance bandwidth of YZY phase-matching (approximately $30^\circ$ C $\times$ cm) yielded good overlap with the ZZZ and ZYY interactions, as can be seen from Fig.~\ref{tripconctempdata}.

\begin{figure}[htb]
\begin{center}
\includegraphics[width=3.25in]{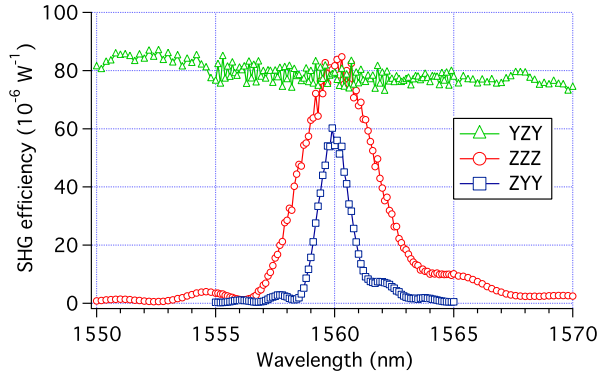}
\end{center}
\vskip -.2in
\caption{Triply concurrent SHG at $37^\circ$C in the $46.3\ \mu$m-period YZY channel.}
\label{tripconcdata}
\end{figure}

Figure \ref{tripconcdata} shows results obtained using the YZY channel with the  $46.3\ \mu$m period, at a temperature of $37^\circ$C. 
%------------
%------------
A fit of the data shows the ZYY peak to occur at exactly 1560 nm at this temperature, while the ZZZ peak occurs at 1560.2 nm. The location of the beam waist in the crystal was adjusted so that the YZY SHG output matched that of  ZZZ. When the waist was moved to maximize YZY, the YZY near-peak efficiency (at $15^\circ$C, see 
Fig.~\ref{tripconctempdata}) became approximately double that of the peak ZZZ efficiency at $37^\circ$C. Using the aforementioned Fourier coefficients and the values $d_{33}=15.4$ pm/V and $d_{32}=d_{24}=3.75$ pm/V \cite{Pack2004}, we obtain a ZYY to ZZZ peak-efficiency ratio of $[(3.75\times 0.3855)/(15.4\times 0.112)]^2 = 2.09/2.97=0.70$, consistent with the experimental results of Figs.~\ref{tripconctempdata},\ref{tripconcdata}. For the YZY to ZZZ peak-efficiency ratio, we obtain $[(3.75 \times 2/\pi )/(15.4\times 0.112)]^2=5.70/2.97=1.92$, again consistent with our experiment. This therefore confirms the values of Ref.~\citenum{Pack2004}.  Note that the initial design used the different values $d_{33}=13.7$ pm/V and $d_{32}=5$ pm/V, which is why the ZYY interaction ends up weaker than ZZZ, but this can clearly be remedied. 

In conclusion, we have designed and experimentally demonstrated a PPKTP crystal with three concurrent phase-matchings at 1560 nm.  The knowledge gained about the YZY QPM period in this work can now be applied to generating a single Fourier-engineered grating for all three processes \cite{Lifshitz2005}. Having a triply concurrent crystal made with a single Fourier-engineered grating gives several advantages over a crystal containing three separate polings.  In particular, the single grating would allow the crystal to be used in single-pass operations, such as those using a nonlinear waveguide, rather than an optical cavity.  For example, just using the simultaneous ZZZ and ZYY phase-matchings in the Fourier-engineered crystal of this work could yield a useful source of collinear polarization-entangled photon pairs. Note that this method could also be used  to make a crystal with four concurrent phase matchings in other materials, such as $\rm LiNbO_3$ and $\rm LiTaO_3$.  Last but not least, and most importantly here, the crystal in this study represents the key component in the implementation of quantum computing over the optical frequency comb \cite{Menicucci2008,Flammia2009}, which is, in theory, extremely scalable. 

MP, PP, and OP were supported by U.S. National Science Foundation grants Nos.\ PHY-0555522, CCF-062210, and PHY-0855632. AA was supported by the Israel Science Foundation, grant no. 960/05 and by the Israeli Ministry of Science, Culture and Sport.

\bibliographystyle{ol}

\end{document}